\documentclass[aps,prb,twocolumn,showpacs,groupedaddress]{revtex4}
\usepackage[american]{babel}
\usepackage{color,graphicx,amsmath,amssymb,bm}

\newcommand\be{\begin{equation}}
\newcommand\ee{\end{equation}}
\newcommand\bea{\begin{eqnarray}}
\newcommand\eea{\end{eqnarray}}

\newcommand\ket[1]{\left|#1\right\rangle}
\newcommand\bra[1]{\left\langle#1\right|}

\newcommand\Avg[1]{\langle#1\rangle}

\newcommand\ra{\rightarrow}

\newcommand\trm{\textrm}

\newcommand\eq[1]{Eq.~(\ref{#1})}

\newcommand\fig[1]{Fig.~\ref{#1}}

\begin{document}

\title{Decoherence of Rabi oscillations of electronic spin states in a double quantum dot}     
  
\author{Alessandro Romito}
\affiliation{
  Department of Condensed Matter Physics, 
  The Weizmann Institute of Science,     
  Rehovot 76100, Israel}

\author{Yuval Gefen}
\affiliation{
  Department of Condensed Matter Physics, 
  The Weizmann Institute of Science,     
  Rehovot 76100, Israel}

\date{\today}

\begin{abstract}
We study the role of charge fluctuations in the decoherence of Rabi oscillations between 
spin states $\ket{\uparrow \downarrow}$, $\ket{\downarrow \uparrow}$ of two electrons in a double dot structure. We consider the effects of fluctuations in energy and in the quantum state of the system, both in the classical and quantum limit. The role of state fluctuations is shown to be of leading order at sufficiently high temperature, applicable to actual experiments. At low temperature the low frequency energy fluctuations are the only dominant contribution.
\end{abstract}

\maketitle

\section{Introduction}

The keystone of quantum information processing is the coherent dynamics of the quantum logical bits (qubits)~\cite{nielsen00}. Although such coherent behavior is well established in atomic systems, it can be maintained only for very short time scales, of the order of few nanoseconds, in charge based solid state based systems. 
To overcome this problem one may employ the spin degree of fredom of the electrons residing in a quantum dot as a qubit~\cite{loss98}. 
In fact, as a consequence of the confined geometry, the coherence  time of the spin may be extended to be of the order of tens of microseconds~\cite{petta05}, primarily restricted by the coupling of the nuclear spin environment via the hyperfine interaction~\cite{nuclei-theory}.
These results motivated the experimental progress in controlling electronic spin in GaAs gated quantum dots systems~\cite{experiments,petta05}. 

In a recent experiment~\cite{petta05} the use of spin states of two electrons in a double dot as the holder of quantum information has been investigated. In that configuration the system is governed by (i) the hyperfine interaction which tends to mix singlet and triplet states and (ii) the exchange interaction which tends to preserve the total spin of the electron pair.
The interplay between the two effects has been studied theoretically~\cite{coish05} and analyzed experimentally~\cite{interplay-exp}.
In particular Ref.~\onlinecite{petta05} reports Rabi oscillations between spin states driven (electrostatically) by tuning the exchange energy. 
Such oscillations (faster than the typical spin decoherence time)
are mainly hindered by charge fluctuations~\cite{burkard99,hu06,klauser06}. 

In this paper we analyze the decoherence effects in the Rabi oscillations due to charge fluctuations. We consider both the effects of exchange {\em energy} fluctuations and fluctuations of the singlet hybridized {\em state} which is affected by charge fluctuations as well.
In particular we calculate the time dependence of the Rabi oscillations  in the presence of gate voltage and tunneling amplitude fluctuations, both in the classical (high temperature, \eq{eq5}) and the quantum case (\eq{eq6}). We describe the crossover of the decoherence rate between low and high temperature regimes, which can be relevant in the actual experiments.
Classical energy fluctuations have been analyzed in Ref.~\onlinecite{hu06}.

\section{The model} 

\begin{figure}[hb]
\begin{center}
\includegraphics[width=70mm]{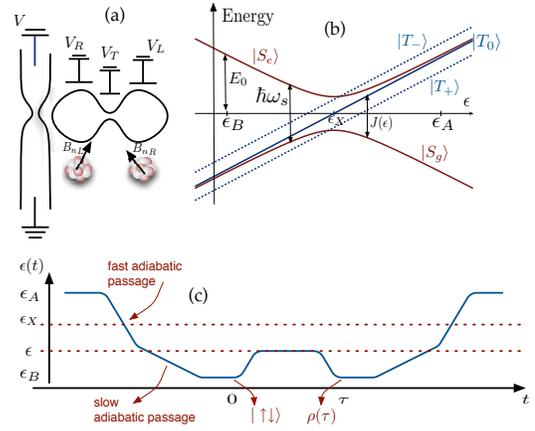}
\end{center}
\caption{(Color online). (a) The schematics of a double dot with a nearby QPC; ${B_N}_L$, ${B_N}_R$ are the respective nuclear magnetic fields. 
(b) Energy levels of lowest singlet and triplet states vs. the detuning parameter $\epsilon$ (c.f. Ref.~\onlinecite{petta05}). 
(c) The time variation of $\epsilon$ needed to drive Rabi oscillations between $\ket{\uparrow \downarrow}$ and $\ket{\downarrow \uparrow}$ states (refer to panel (b)). 
The system is initially in its ground state $\ket{S_g (\epsilon_A)}$. $\epsilon$ is then varied adiabatically ($\hbar \dot{\epsilon} \ll J(0)^2/E_0 $) to $\epsilon=\epsilon_B$,  keeping the system in its lowest $S_z=0$ state, either $\ket{\uparrow \downarrow}$ or $\ket{\uparrow \downarrow}$. 
During this variation the point $\epsilon=\epsilon_X$ is crossed at a time faster than $ \hbar/(g \mu_B \max\{{B_N}_L,{B_N}_R\})$ to avoid  a transition to $\ket{T_+}$ (but with $\hbar \dot{\epsilon} \ll J(0)^2/E_0$); approaching  $\epsilon=\epsilon_B$, instead, the variation of $\epsilon$ is slowed down ($\hbar E_0 \dot{\epsilon} \ll \Avg{T_0|H_N|S_g}^2 \ll J(0)^2$) to guarantee adiabaticity with respect to the nuclear interaction. 
Following the pulse that induces Rabi oscillations in the interval $[0, \tau]$, the adiabatic variation of $\epsilon$ at $t>\tau$ is reversed. 
This allows to identify the state of the system at $t=\tau$ by mapping the states $\ket{\uparrow \downarrow}$  and $\ket{\downarrow \uparrow}$ at $\epsilon=\epsilon_B$ to $\ket{S_g(\epsilon_A)}$ and $\ket{T_0}$ at $\epsilon=\epsilon_A$ respectively, where the latter can be measured employing the QPC.
}
\label{schema}
\end{figure}

The system (Cf. Petta {\it et} al.~\cite{petta05}) is schematically presented in \fig{schema}. 
It consists of a gate confined semiconducting double quantum dot. Tunnel barriers connect each dot to the adjacent reservoirs allowing dot-lead tunneling of electrons.
The gate voltages, $V_T$, $V_L$, and $V_R$, control the tunnel between the dots, and the dots' charge configuration $(n_L,n_R)$, respectively.  
It is possible to measure such a charge configuration using a quantum point contact (QPC) located near one the dots.
The dimensionless detuning parameter, $\epsilon \propto V_L-V_R$ 
controls the difference $n_L-n_R$. In Ref.~\onlinecite{petta05} the system was operated between $(1,1)$ and $(0,2)$. 
In the $(0,2)$ charge configuration ($\epsilon=\epsilon_A$), the antisymmetric nature of the electron wave function enforces the ground state of the system to be a singlet. The excitation energy to the lowest triplet state is experimentally estimated to be $\gtrsim 400 \mu eV $, larger than the charging energy of the single $(1,1)$ state, $E_0$. In the $(1,1)$ configuration ($\epsilon=\epsilon_B$) the singlet and triplet states are instead practically degenerate.
The low energy states of the system are two singlets, $\ket{S_L}, \ket{S_R}$, 
corresponding to $(1,1)$ and $(0,2)$ respectively, and the three triplet states for charge configuration $(1,1)$, $\ket{T_0}, \ket{T+}, \ket{T_-}$, respectively with the spin component $0,1,-1$ in the $\hat{\mathbf{z}}$ direction perpendicular to the dots' plane.
An external magnetic field ${\bf B} = B \hat{{\bf z}}$ is applied to
 split the states $\ket{T_+}, \ket{T_-}$ by the Zeeman energy $\Delta_z= g\mu_B B \sim 2.5 \mu eV$. 
 We neglect the role of these states (see below) and we write  
the Hamiltonian of the system in the $S_z=0$ subspace as
\bea
\hat{H_0} & = & E_0 \epsilon (\ket{T_0}\bra{T_0} + \ket{S_L}\bra{S_L} 
 - \ket{S_R}\bra{S_R} ) \nonumber \\ 
 & &+ E_0( \lambda_s \ket{S_L}\bra{S_R}  + \trm{h.c.} ) \, .
\label{eq1}
\eea
We introduced a tunneling amplitude between the two singlet states, $\lambda_s$, which is the only possible tunneling matrix element assuming conservation of total spin. 
It can be chosen real and positive. 
This leads to hybridization of the ground and excited states in the singlet subspace,
\bea
  \ket{S_g(\epsilon)} &=& -\sin \theta \ket{S_R}+ \cos \theta \ket{S_L} \, , \nonumber \\
\ket{S_e(\epsilon)} &=& \cos \theta \ket{S_R}+ \sin \theta \ket{S_L} \, , 
\eea  
thence 
\be
\hat{H_0}=\frac{\hbar \omega_s}{2}(\ket{S_e(\epsilon)}\bra{S_e(\epsilon)}-\ket{S_e(\epsilon)}\bra{S_e(\epsilon)}) + E_0\epsilon \ket{T_0}\bra{T_0} \, , \nonumber
\ee 
with $ \hbar \omega_s =2 E_0 \sqrt{\epsilon^2 + \lambda_s^2}$, $\tan \theta = (\sqrt{\epsilon^2+\lambda_s^2} + \epsilon)/\lambda_s $.
The energy levels as a function of $\epsilon$ are plotted in \fig{schema}. 

Transitions between singlet and triplet states are made possible due to the hyperfine interaction of the electrons in the dot with nuclear spins, which can be written in terms of the effective nuclear magnetic field in each of the dot, $H_N=g \mu_B ({{\bf B}_N}_L \cdot {\bf S_L} + {{\bf B}_N}_R \cdot {\bf S_R})$. 
The typical dynamical scale of the nuclear environment is of the order of tens of microseconds, and therefore it acts as if it is a frozen external field over the duration of the experiment. 
With $ {B_N}_L, {B_N}_R \sim 5 mT \ll B$, the hyperfine interaction is 
effective only at $\epsilon \sim \epsilon_X$, where it can mix $\ket{S_g}$ and $\ket{T_+}$, and around $\epsilon \sim \epsilon_B \ll 1$ where it mixes the lower energy states $\ket{S_g}$ and $\ket{T_0}$: $H_N= g \mu_B ({{\bf B}_N}_L - {{\bf B}_N}_R)\cdot {\bf z}\ket{T_0}\bra{S_g}+\trm{h.c.}$, while the energy difference between them is $J(\epsilon) = E_0(\epsilon +\sqrt{\epsilon^2+\lambda_s^2}) \ll \Avg{T_0|H_N|S_g}$. The ground state of the system at $\epsilon=\epsilon_B$ is therefore $(\ket{S_g(\epsilon=-1)} \pm \ket{T_0})/\sqrt{2}=\ket{\uparrow \downarrow}$ (for $+$) or $\ket{\uparrow \downarrow}$. $\ket{\uparrow \downarrow}$ or $\ket{ \uparrow \downarrow}$ ($-$): the spin in the two dots are oppositely oriented. Hereafter we consider $\ket{\uparrow \downarrow}$ state.

In the experiment described in Ref.~\onlinecite{petta05} the detuning parameter is varied in time to induce Rabi oscillations between $\ket{\uparrow \downarrow}$ and $\ket{\uparrow \downarrow}$.
The time dependence of the parameter $\epsilon$ used to drive the oscillations is depicted in \fig{schema}.
The system is prepared in the state $\ket{\uparrow \downarrow}$ (or equivalently $\ket{\downarrow\uparrow} $) at $\epsilon=\epsilon_B$ (cf. \fig{schema}).
Subsequently, a gate voltage pulse at $t=0$ modifies $\epsilon$ to a point where $\Avg{T_0|H_N|S_g} \ll J(\epsilon)$, thus  inducing oscillations between $\ket{\uparrow \downarrow}$ and $\ket{\downarrow\uparrow} $ over a time interval $\tau$. 
The following manipulation of $\epsilon$ (cf. \fig{schema}) allows to relate the measured conductance of the QPC with the probability of finding the system in $\ket{\uparrow \downarrow}$ right after the pulse, $P(\tau)=|\bra{\uparrow \downarrow}\exp(-i \hat{H}_0 \tau) \ket{\uparrow \downarrow}|^2$.

\section{Classical noise}. 

The Rabi oscillations are obtained by tuning the energy difference $J(\epsilon)$ between $\ket{S_g}$ and $\ket{T_0}$, which results in the different charge of the triplet and hybridized singlet. 
Rabi oscillations will therefore be extremely sensitive to an environment coupled to charge as opposed to the nuclear spin environment. 
Decoherence effects will originate both from fluctuations of the (exchange) energy $J(\epsilon)$, analyzed in Ref.~\onlinecite{hu06}, and fluctuations of the hybridized singlet state $\ket{S_g(\epsilon)}$.
We analyze the Rabi oscillations taking into account fluctuations of $V_L$, $V_R$ and $V_T$. The respective gates are controlled independently of each other, hence it is natural to assume that their fluctuations are independent.
In principle it is possible to determine a correlation matrix for the fluctuations of the parameters in the Hamiltonian, $\epsilon$ and $\lambda_s$, by considering a specific potential form for the double dot. Instead we assume that $V_T$ affects only the tunneling matrix elements, $\lambda_s$, which is reasonable for weak tunneling. Then $\epsilon$ is affected only by the fluctuations of $V_L-V_R$.
The Hamiltonian is
\bea
\hat{H} &=& \hat{H}_0(\{ \epsilon \ra \epsilon +\xi_c(t) ,\,  \lambda_s \ra \lambda_s +\xi_{\lambda} (t) \}) \nonumber \\
&=& \hat{H}_0 + E_0(\hat{V}_c \xi_c(t) + \hat{V}_{\lambda} \xi_{\lambda} (t) ) \, ,
\label{eq2}
\eea
where $\xi_i(t)$ are Gaussian distributed with $\Avg{\xi_i(t)}=0$, $\Avg{\xi_i(t) \xi_j(t')}= 2 \hbar^2 \Gamma_i/E_0^2 \delta(t-t') \delta_{i,j}$ for $i,j \in \{ c,s \}$ and
$\hat{V}_c=\ket{T_0}\bra{T_0}+\ket{S_L}\bra{S_L}-\ket{S_R}\bra{S_R}$, $\hat{V}_{\lambda}=\ket{S_L}\bra{S_R}+\ket{S_R}\bra{S_L}$. 
The assumption of white noise renders the dynamics Markovian,  leading to an exact master equation for the density matrix of the double dot,
\be
\partial_t \rho = -i/\hbar [\hat{H_0},\rho]-\sum_{j \in \{c,\lambda\} } \Gamma_j 
(\hat{V}_j^2 \rho -2 \hat{V}_j \rho \hat{V}_j +\rho \hat{V}_J^2) \, .
\label{kinetic_eq}
\ee
This differential equation is solved with the initial condition $\rho=\ket{\uparrow \downarrow}\bra{\uparrow \downarrow}$ by neglecting terms of order $\mathcal{O}(J(\epsilon)/(\hbar \omega_s)=\sin^2\theta)$.  
Under this assumption the density matrix at any time $t>0$ can be written as 
\bea
\rho (t) &=& \frac{1}{2} \left[ \ket{T_0}\bra{T_0}+ Y(t) \ket{S_g}\bra{S_g} + (1-Y(t)) \ket{S_e}\bra{S_e} \right.\nonumber \\ 
&& + \left.(X(t)\ket{S_g}\bra{T_0}+\trm{h.c.})\right] \, ,
\label{eq4}
\eea 
in terms of the two functions $Y(t)$ and $X(t)$ describing the evolution of the state populations and coherency respectively. The explicit expressions, $X(t)=e^{(i J(\epsilon)/\hbar-\gamma_2) t}$, $Y(t)=(1+e^{-\gamma_1t})/2$, allow to determine the surviving probability $P(t)=\bra{\uparrow \downarrow} \rho(t) \ket{\uparrow \downarrow}=1/4[1+Y(t)+2 \Re e \{ X(t)\}]$,  
\bea
&  P(\tau) = \frac{1}{8} \left[3 + e^{-\gamma_1 \tau}+ 4\cos(J(\epsilon) \tau/\hbar)e^{- \gamma_2 \tau}\right] \, , & \label{eq5}  \\
&  \begin{array}{l}
\gamma_1=4(\Gamma_c \sin^2(2\theta) +\Gamma_{\lambda} \cos^2(2 \theta)) \, , \vspace{0.15cm} \\
 \gamma_2=4\Gamma_c \sin^2( \theta) + \Gamma_{\lambda} \, . 
 \end{array} &  \label{classical noise} \\
& \trm{- {\em classical}-} & \nonumber
\eea 

The measured probability consists of damped oscillations around a 
mean that approaches an asymptotic value. 
The decay of the oscillations is related only to the decoherence of $\rho(t)$, while the relaxation of the populations, encoded in $Y(t)$, determines the slow time variation of the mean. 
The dependence of the decay rates, $\gamma_1$, $\gamma_2$, on $\epsilon$ is quite different from what we would have obtained by simply accounting for fluctuation of $J(\epsilon)$, in which case  
\bea
&  
\gamma_1=0 \, , \gamma_2=4\Gamma_c\sin^4\theta +\Gamma_{\lambda}\sin^2(2\theta) \, .
&  \label{classical noise energy only} \\
&  \trm{- {\em classical energy fluctuations only} -} & \nonumber \, 
\eea
This means that the effects of fluctuations of the state $\ket{S_g}$ 
cannot be neglected with respect to (exchange) energy fluctuations.
We also note that the asymptotic value of the probability is $P(\tau) = 3/8$, corresponding to a steady state density matrix with equally populated singlet states ($\ket{S_g}$, $\ket{S_e}$). 
This is a signature of the high temperature ($k_B T \gg \hbar \omega_s$) limit, and is related to the assumption of classical gate voltage fluctuations. 
By contrast, at low temperature we expect that only the lower singlet level is populated. 
In the experiment oscillations with $J(\epsilon)$ ranging from tenths to few $\mu eV$ have been observed  at $k_BT \sim 10 \mu eV$~\cite{petta05}. 
We therefore expect that the condition $k_B T \sim \hbar \omega_s \gg J(\epsilon)$ is that obtained experimentally, in which case quantum state fluctuations are important and we explore their dependence on the temperature.

\section{Quantum noise}. 

In order to extend our analysis beyond the high temperature limit we need to consider the quantum nature of the gate voltage fluctuations. We therefore modify our classical model (\eq{eq2}):
\be
\hat{H} = \hat{H}_0 + \hat{V}_c A_c +  \hat{V}_{\lambda} A_{\lambda} +{H_{\trm{bath}}}_{c} +{H_{\trm{bath}}}_{\lambda} \, .
\ee
Here the operators $\hat{V}_{c(\lambda)}$ acting on the system are the same as in \eq{eq2} while 
the classical fluctuators are replaced by the operators $A_{c(\lambda)}=\sum_i {a_{c(\lambda)}}_i ({b_{c(\lambda)}}^{\dag}_i+{b_{c(\lambda)}}_i)$ of bosonic baths, ${H_{\trm{bath}}}_{c(\lambda)}=\sum_i \hbar \omega_i {b_{c(\lambda)}}^{\dag}_i{b_{c(\lambda)}}_i$. 
The effect of the reservoirs on the dynamics of the electrons 
in the double dot is entirely characterized by their symmetric and antisymmetric spectral functions 
\be
S^{\pm}_{c(s)}(\omega)=1/(2\pi) \int_{\mathbb{R}} dt \, e^{i \omega t}\Avg{[A_{c(\lambda)}(t),A_{c(\lambda)}(0)]_{\pm}} \, .
\ee
We take both baths to be at equilibrium at the same temperature $k_B T=1/\beta$.
Their bosonic nature guarantees that $S^{+}(\omega)=\coth(\beta \hbar \omega/2)S^{-}(\omega)$, where $S^{-}(\omega)$ is temperature independent, $S^{-}_{c(\lambda)}(\omega)=\sum_i {a_{c(\lambda)}}_i^2 [\delta(\omega+\omega_i)-\delta(\omega-\omega_i)]$.

We assume that the bath is weakly coupled to the system, and we determine the evolution of the density matrix to second order in ${a_{c(\lambda)}}_i/\omega_s$. 
Following the Bloch-Redfield approximation~\cite{cohen92}, we introduce a (short) bath correlation time, $\bar{\tau}$, characterizing the typical time scale at which any 
correlation of the system and the reservoir disappears. 
The time evolution of the reduced density matrix of the system, $\rho$, coarse grained at time scales
$\Delta t \gg \bar{\tau}$, is Markovian. It is determined by the first order linear differential equation 
\be
\partial_t \rho_{a,b}=-i \omega_{a,b} \, \rho_{a,b}- \sum_{c,d}\mathcal{R}_{a,b,c,d} \,  \rho_{c,d} e^{-i(\omega_{a,b}-\omega_{c,d})}\, ,
\label{evolution}
\ee
written in the basis of eigenstates of $\hat{H}_0$ where $\hbar \omega_{a,b}=\Avg{a|\hat{H}_0|a}-\Avg{b|\hat{H}_0|b}$ and $\mathcal{R}_{a,b,c,d}$ is the Bloch-Redfield tensor:
\bea
& &\mathcal{R}_{a,b,c,d}=\sum_{j=c,\lambda}\left[ \sum_n \delta_{b,d} \Avg{a|V_j|n}\Avg{n|V_j|c}g_j(\omega_{c,n}) \right.  \nonumber \\
& & - \Avg{a|V_j|c}\Avg{d|V_j|b}g_j(\omega_{c,a}) - \Avg{a|V_j|c}\Avg{d|V_j|b}g_j(\omega_{b,d})  \nonumber \\
& & \left. + \sum_n \delta_{a,c} \Avg{d|V_j|n}\Avg{n|V_j|b}g_j(\omega_{n,d}) \right] \, ,
\eea
with $g(\omega)=1/4[S^+_j(\omega)+S^-_j(\omega)]-i \int_{\mathbb{R}} P dx/(2\pi) \, (S^+_j(x)+S^-_j(x))/(x-\omega)$.
At times $t \gtrsim \hbar/J(\epsilon) \gg \hbar/\omega_s$, neglecting terms of order $\mathcal{O}(J/\omega_s)$, the sum in \eq{evolution} involves only terms such that $\hbar (\omega_{a,b}-\omega_{c,d}) \ll 0$. Explicitly the only relevant entries of the Bloch-Redfield tensor are:
$\mathcal{R}_{S_g,S_g,S_g,S_g}=-\mathcal{R}_{S_e,S_e,S_g,S_g}$,
$\mathcal{R}_{S_e,S_e,S_e,S_e}=-\mathcal{R}_{S_g,S_g,S_e,S_e}$,
$\mathcal{R}_{S_g,T_L,S_g,T_L}=\mathcal{R}^{*}_{T_L,S_g,T_LS_g}$.
It follows that state population and coherency evolve independently 
of each other and therefore the density matrix has the same form presented in \eq{eq4}, with different functions $X(t)$ and $Y(t)$.

Once the expression of $\rho(t)$ is known, it can be used to calculate the survival probability, 
\bea
& & P_{qm}(\tau) = \frac{1}{8} \big[ 3+ e^{-\gamma_1 \tau}+ \tanh\left( \frac{\hbar \beta \omega_s}{2}\right)(1-e^{-\gamma_1 \tau}) \nonumber \\
 & &  \hspace{2cm} + 4\cos((J(\bar{\epsilon})/\hbar+\Delta_J)\tau)e^{-\gamma_2 \tau} \big] \, , \label{eq6}  \\
& & \gamma_1 = 2 \pi  [\sin^2(2\theta)  S_c^+(\omega_s)+ \cos^2(2\theta)   S_{\lambda}^+(\omega_s)]/\hbar^2  \, ,  \\  
& & \gamma_2 =\pi  \left[ \sin^2(2\theta) (S_c^+(\omega_s)-S_c^-(\omega_s)) + \cos^2(2 \theta) \right. \nonumber \\
& &  \hspace{1cm} (S_{\lambda}^+(\omega_s)-S_{\lambda}^-(\omega_s)) +  (\cos(2 \theta)-1)^2  S_c^+(0) \nonumber \\
& & \hspace{1cm} \left. +\sin^2(2\theta) S_{\lambda}^+(0) \right]/(2\hbar^2) \, , \label{quantum}
\eea
and
$\hbar^2 \Delta_J=[\sin^2(2 \theta)\int_{\mathbb{R}} P \,d \omega /\omega \, ( S_{\lambda}^{-}(\omega)-  S_c^{-}(\omega))+\int_{\mathbb{R}} P  d \omega/(\omega+\omega_s) \, (\sin^2(2\theta ) (S_c^+(\omega)+S_c^-(\omega)) 
+\cos^2(2\theta)(S_{\lambda}^+(\omega)+S_{\lambda}^-(\omega)))]/2$.
$\Delta_J$ is a shift in the frequency of the Rabi oscillations that can be neglected compared with $J(\epsilon)/\hbar$, 
consistent with our second order perturbation expansion. 
The Bloch-Redfield approximation employed implies that \eq{eq6} is valid in the limit $\gamma_{1(2)} \ll \omega_s,  1/\bar{\tau}$.  We note that, unlike $\gamma_2$,
$\gamma_1$ depends only on the symmetric (classical) correlators, $S_{c(\lambda)}^+(\omega)$. $\gamma_2$ consists of contributions from the bath correlation function at frequency $\omega_s$ (which describes the relaxation process between the two singlet eigenstates and the corresponding contribution to the dephasing), and from the zero frequency correlation function (corresponding to the contribution of pure dephasing)~\cite{whitney04}. 

\begin{figure}
\begin{center}
\includegraphics[width=70mm]{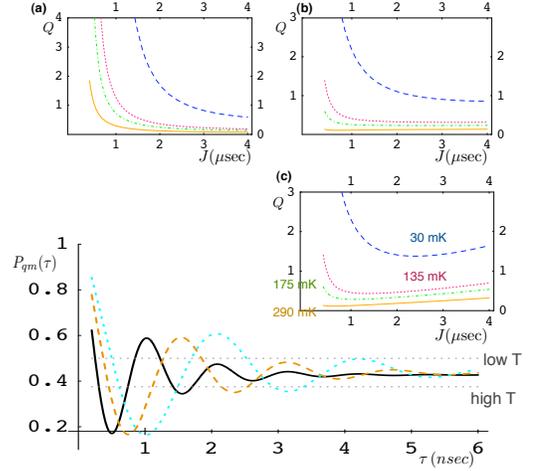}
\end{center}
\caption{(color online) Surviving probability versus time $\tau$ (\eq{eq6}) at $J(\epsilon)=3.9 \, \mu\trm{eV}$ (black solid line), $J(\epsilon)=2.4 \, \mu\trm{eV}$ (orange dashed line) and $J(\epsilon)=2 \, \mu \trm{eV}$ (cyan dotted line). For all the curves $J(0)=5$  $\mu\textrm{eV}$, $T=135$ $\textrm{meV}$ as estimated from the experiment~\cite{petta05}, and $\alpha_c=\alpha_{\lambda}=7 \times 10^{-3}$ obtained by fitting with the experimental curves. The high and low temperature asymptotic values are depicted (dotted gray lines). Insets: $Q=J/(2 \pi \hbar \gamma_2) \propto \, \trm{(number of visible oscillations)}$ as a function of $J(\epsilon)$ for three possible scenarios: (a) $ \alpha_c=0.02$, $\alpha_{\lambda}=0$; (b) $\alpha_c=\alpha_{\lambda}= 7 \times 10^{-3}$; (c) $\alpha_c=0$, $\alpha_{\lambda}=7 \times 10^{-3}$. Different curves correspond to different temperatures according to their labels in the inset (c); in all plots $J(0)=5 \, \mu \trm{eV}$.
}
\label{risultati}
\end{figure}

In the high temperature limit, $\beta \hbar \omega_s \ra 0$, $S^-_{c(\lambda)}(\omega_s)$ is negligibly small as compared with $S^+_{c(\lambda)}(\omega_s)$, we thus expect a classical result.  
If we furthermore assume a Ohmic bath, i.e. $S^-_{c(\lambda)}(\omega)=\alpha_{c(\lambda)} \hbar^2 \omega$, 
such that at high temperature $S^+_{c(\lambda)}(\omega) \sim S^+_{c(\lambda)}(0)$ 
for $0<\omega<\omega_s$, 
\eq{eq6} reduces to \eq{eq5} with
\bea
& \begin{array}{l} 
\gamma_1=2 \pi / \hbar (S^+_c(0) \sin^2(2 \theta) + S^+_{\lambda}(0)\cos^2(2 \theta))\, , 
\vspace{0.15cm} \\ 
\gamma_2=\pi/(2\hbar)(4S^+_c(0) \sin^2(\theta)S^+_{\lambda}(0)) \, .
\end{array} & \label{quantal high T} \\
& \trm{- {\em quantal high $T$} -} & \nonumber
\eea
We in fact recover the classical result for white noise fluctuations (\eq{classical noise}) by identifying 
$\Gamma_{c(\lambda)}=\pi/(2 \hbar^2) S_{c(\lambda)}^+(\omega_s)$.
At low temperature, $\beta \hbar\omega_s \gg 1$, $S^+_{c(\lambda)}(\omega_s)\simeq S^-_{c(\lambda)}(\omega_s)$ and  the quantum nature of the bath becomes important. 
The rate $\gamma_1$ disappears from the expression for $P(t)$ and we also note the finite frequency contribution to $\gamma_2$ vanishes.
Only the zero frequency component of the spectral density of the bath (which is responsible for pure dephasing) survives,
\bea
&  \gamma_2= \pi/(2 \hbar^2) ( 4 S_{c}^+(0) \sin^4\theta+ S_{\lambda}^+(0) \sin^2(2 \theta)) \, .  \\
&  \trm{- {\em quantal low $T$} -} & \nonumber  
\label{quantal low T}
\eea
In this limit the dependence of $\gamma_2$ on $\theta$ can be explained entirely in terms of classical fluctuations of the oscillation frequency, $J(\epsilon)$ (cf. \eq{classical noise energy only}). 
Indeed, the effects of fluctuations of the state $\ket{S_g}$ do involve transitions between the latter state and the singlet excited state, yet these transitions are exponentially suppressed by $e^{- \beta \omega_s}$. Note however that experimentally the regime $\beta \hbar \omega_s \lesssim 1$ can be reached, hence the effect of fluctuations of the {\em state} $\ket{S_g}$ can be of interest.
In particular this fluctuations affect the steady state value of the survival probability which is a function of $\beta \omega_s$, $P_{qm}(\tau \ra \infty) = 1/8[3+\tanh(\beta \hbar \omega_s/2)]$ (cf. \fig{risultati}) and can be directly observed in the experiments. 

A comparison with the experimental results of Ref.~\onlinecite{petta05} is obtained assuming Ohmic baths with spectral densities $S^-_{c(\lambda)}(\omega)=\alpha_{c(\lambda)}\hbar^2 \omega$. These properly describe charge fluctuations due to the external circuit controlling the gate voltages.
An analysis of the possible scenarios $\alpha_c \gtreqless \alpha_{\lambda}$, shows that the experimental fact that the number of visible oscillations as function of $J(\epsilon)$ is constant is correctly reproduced for $\alpha_c  \lesssim \alpha_{\lambda}$ (cf. insets in \fig{risultati}).
The time dependence of $P_{qm}(\tau)$ is depicted in \fig{risultati} for different values of $J(\epsilon)$. We obtain a fit with the experimental data for $\alpha_c=\alpha_{\lambda}=7 \times 10^{-3}$, consistent with the strength of electromagnetic environment in other solid state systems~\cite{astafiev04}.

\section{Extended model}

The previous analysis is now extended to include the lowest energy triplet state in the charge configuration $(0,2)$, $\ket{T_0'}$. 
This might be necessary if the energy of $\ket{T_0'}$ is comparable with the Coulomb energy $E_0$. 
The new Hamiltonian reads, $\hat{H}_0'=\hat{H}_0+E_0(\lambda_t \ket{T_0}\bra{T_0'}+\trm{h.c.})+E_0(\delta- \lambda_s \epsilon)\ket{T_0'}\bra{T_0'}$, where 
$\delta$ is the excitation energy to the triplet state in the $(0,2)$ configuration.
Owing to electron tunneling, $\lambda_t$, the two triplet states, $\ket{T_0}$ and $\ket{T_0'}$, do hybridize (cf. \fig{schema}(c))~\cite{stopa06}.  
The energy spectrum of singlet and triplet states in the subspace $S_z=0$ is depicted in \fig{schema}(c). 
The energies of the hybridized triplet states, $\ket{T_g(\epsilon)} = -\sin \varphi \ket{T_0'}+ \cos \varphi \ket{T_0}$ and $\ket{T_e(\epsilon)} = \cos \varphi \ket{T_0'}+ \sin \varphi \ket{T_0}$, are $E_0 \delta \mp \hbar \omega_t/2=E_0\delta \mp E_0 \sqrt{(\epsilon-\delta)^2+\lambda_t^2}]$ respectively, with $\tan \varphi = (\sqrt{(\epsilon-\delta)^2+\lambda_t^2} + \epsilon-\delta)/\lambda_t$. 
The Hamiltonian for this model includes four parameters, $\epsilon, \delta, \lambda_s, \lambda_t$ which depend on three fluctuating gate voltages only, $V_L, V_R, V_T$. 
In principle it is possible to determine a correlation matrix for the fluctuations of the parameters in the Hamiltonian by considering a specific potential form for the double dot. 
Instead we assume that $V_T$ affects only the tunneling matrix elements, which is reasonable for weak tunneling. 
The fluctuations of $\lambda_t \ra \lambda_t+\xi_t(t)$ and $\lambda_s \ra \lambda_s +\xi_s(t)$ will then depend on the same bath and will therefore be correlated,  
$\xi_t(t)=f \xi_s(t)$ with $f=(\partial \lambda_t /\partial V_T)(\partial V_T /\partial \lambda_s)$. 
At the same time the gate voltage difference $V_L-V_R$ will affects only $\epsilon$.
The density matrix now evolves as (cf. \eq{eq4}) $\rho'(t)=\rho (t) +1/2 W(t) \ket{T_g}\bra{T_g} + 1/2 (1-W(t))\ket{T_e}\bra{T_e}$,  and the probability of finding the system in the $\ket{\uparrow \downarrow}$ at time $\tau$ is $P_{qm}'(\tau)=P_{qm}(\tau)-1/8(1-\tanh(\beta \hbar \omega_t/2))(1-e^{-\gamma_3 \tau})$, with $\gamma_3 =2 \pi /\hbar^2 [\sin^2(2\varphi)  S_c^+(\omega_t)+ f^2 \cos^2(2\varphi) S_{\lambda}^+(\omega_t)]$ and $\gamma_2$ replaced by $\gamma'_2=  \gamma_1/4 (1-\tanh(\beta \hbar \omega_s)) 
+\gamma_3/4 (1-\tanh(\beta \hbar \omega_t))
+\pi/(2\hbar^2)[(\cos^2(2\theta)-\cos^2(2\varphi))S_c^+(0)
+(\sin^2(2\theta)-f\sin^2(2\varphi))S_{\lambda}^+(0)]$. 
The physical mechanism that induces decoherence in $P(\tau)$ is the same described in the previous paragraphs.
Similarly to the decoherence in the singlet subspace, fluctuations in $\ket{T_g}$ do involve now the exited triplet state $\ket{T_e}$, an effect that is small in $ e^{-\beta \hbar \omega_t/2}$. 
Note that, even at low temperature, $k_BT \ll \hbar\omega_t$, fluctuations in the energy $J(\epsilon)$ (cf. \fig{schema}(c)) modify $P_{qm}(\tau)$. Remarkably in this case, even 
in the presence of a ``sweet point''($\theta=\varphi$)~\cite{stopa06}, while $\partial_{\epsilon} J=0$, fluctuations of the tunneling rates are important due to the difference between electron tunneling in the triplet and singlet states.

\section{Conclusions}

We have presented here a simple model describing the effect of charge fluctuations on Rabi oscillations between spin states $\ket{\uparrow \downarrow}$ and $\ket{\downarrow \uparrow}$ of electrons in a double dot. We have accounted for decoherence effects due to both energy and quantum state fluctuations, by including the quantum effects of a fluctuating environment within the Born-Markov approximation ---Eqs.(\ref{eq6}--\ref{quantum}). We have shown that not only in the high temperature limit does the result reproduce that of classical fluctuations (compare \eq{quantal high T} to \eq{classical noise}), but also the low temperature result has a classical interpretation in terms of energy fluctuations only (not state fluctuations, compare \eq{quantal low T} to \eq{classical noise energy only}).
In fact the role of the state fluctuations is significant at a temperature that exceeds the singlet excitation energy, a regime which is accessible experimentally. Note that at high temperature the ``classical limit'' refers to classical environment induced fluctuations. The latter can still cause fluctuations in the {\em quantum state} of the system. 
At low temperature state fluctuations are frozen out.

We are grateful to Amir Yacoby and Sandra Foletti for useful discussions. 
We acknowledge the support of U.S.-Israel BSF, the ISF of the Israel Academy of Sciences and DFG project SPP 1285.


\end{document}